\documentclass[superscriptaddress,showkeys,reprint,nofootinbib]{revtex4-1}
\usepackage{graphicx}
\usepackage[margin=0.7in]{geometry}
\usepackage{amsmath}
\usepackage{hyperref}
\usepackage{multirow}

\setcitestyle{super}

\renewcommand{\Im}{\operatorname{Im}}
\renewcommand{\Re}{\operatorname{Re}}
\newcommand{\sub}[1]{\ensuremath{_{\textrm{#1}}}}  
\newcommand{\Caltech}{Department of Applied Physics and Materials Science, California Institute of Technology, Pasadena, CA, USA}
\newcommand{\RPIMSE}{Department of Materials Science and Engineering, Rensselaer Polytechnic Institute, Troy, NY, USA}
\newcommand{\HarvardSEAS}{John A. Paulson School of Engineering and Applied Sciences, Harvard University, Cambridge, MA, USA}
\newcommand{\MITPhy}{Department of Physics, Massachusetts Institute of Technology, Cambridge, MA, USA}
\newcommand{\UPenn}{School of Engineering and Applied Sciences, University of Pennsylvania, Philadelphia, PA, USA}
\newcommand{\Zagreb}{Department of Physics, Faculty of Science, University of Zagreb, Zagreb, Croatia}

\usepackage[usenames]{color}

\begin{document}

\title{Ultra-light \AA-scale Optimal Optical Reflectors}

\author{Georgia T. Papadakis}\affiliation{\Caltech}
\author{Prineha Narang}\email{pnarang@fas.harvard.edu}\affiliation{\HarvardSEAS}
\author{Ravishankar Sundararaman}\affiliation{\RPIMSE}
\author{Nicholas Rivera}\affiliation{\MITPhy}
\author{Hrvoje Buljan}\affiliation{\Zagreb}
\author{Nader Engheta}\affiliation{\UPenn}
\author{Marin Solja\v{c}i\'{c}}\affiliation{\MITPhy}

\date{\today}

\begin{abstract}
\textbf{\uppercase{Abstract}}
High-reflectance in many state-of-the-art optical devices is achieved with noble metals. 
However, metals are limited by losses, and for certain applications, by their high mass density.
Using a combination of \emph{ab initio} and optical transfer matrix calculations,
we evaluate the behavior of graphene-based \AA-scale metamaterials and find that
they could act as nearly-perfect reflectors in the mid-long wave infrared (IR) range.
The low density of states for electron-phonon scattering and interband excitations leads
to unprecedented optical properties for graphene heterostructures, especially alternating
atomic layers of graphene and hexagonal boron nitride, at wavelengths greater than $10~\mu$m.
At these wavelengths, these materials exhibit reflectivities exceeding 99.7\%
at a fraction of the weight of noble metals, as well as plasmonic mode confinement
and quality factors that are greater by an order of magnitude compared to noble metals. These findings hold promise for
ultra-compact optical components and waveguides for mid-IR applications.
Moreover, unlike metals, the photonic properties of these heterostructures could be actively tuned via chemical and/or electrostatic doping, providing exciting possibilities for tunable devices.
\end{abstract}

\maketitle

Two-dimensional (2D) materials exhibit a diverse array of electronic,
photonic and phononic properties,\cite{RevModPhys.81.109,
Novoselov666, Geim:2007ty, R.:2010qv,PhysRevB.76.073103, Vakil1291}
while stacked 2D materials, or van der Waals (vdW) heterostructures
further expand the scope for engineering new material properties by
combining 2D layers.\cite{Xia:2014lq, Geim:2013ud, Jariwala:2016zp}
Heterostructures involving graphene with other 2D materials have emerged
as a class of materials that demonstrate strong interaction with light,
making it possible to realize a variety of new optical phenomena
and nanophotonic devices, covering spectral ranges from the microwave
to the ultraviolet.\cite{Ci:2010rz,PhysRevB.80.245435, Bonaccorso:2010uo}
With the rapid advances in fabrication of two-dimensional (2D) materials,\cite{Butler:2013fv}
the possibility of building electronic circuits\cite{Akinwande:2014zl} and
photonic devices\cite{DaiS.:2015sf, Caldwell:2016rm} that harness the mixed
functionalities of vdW heterostructure materials is now a reality. 

Nearly perfect reflection is a requirement in designing 
compact waveguides, engineering emission for back-reflectors
in solar energy technologies,\cite{Kim:2015nr, PhysRevB.90.165409}
as well as for macroscopic objects like aircrafts.
So far, aside from the Bragg mirrors which are limited by narrow bandwidth, the two most widely used materials for mirroring systems in the mid-long wave IR are gold and silver, primarily due to their excellent electric conductance in the IR. 
Increased ohmic losses together with high mass density, which is crucial for state-of-the-art aerospace technology, limit the performance of noble metals. 
Here, we envision engineering a new class of ultra-reflective materials that are also ultra-light, by taking advantage of the unique properties of graphene and its vdW heterostructures.

Critical parameters for a good electric conductor are the relaxation time,
effective mass and number density of the charge carriers (electrons).
In conventional noble metals, the high density of states near the Fermi level leads to
small relaxation times, and the effective mass is approximately the free electron mass.
The two-dimensional nature of graphene and its unique linear electronic dispersion
yield a much larger relaxation time due to the lower electronic density of states,
as well as an effective mass approaching zero, both of which enhance
the carrier mobility in comparison to metals.
Although graphene typically has lower carrier density than Ag and Au, its important features of 
high mobility and tunable Fermi level (through external bias or carrier injection)
motivates us to seek graphene-based heterostructures that might surpass
the photonic and plasmonic properties of noble metals.

\begin{figure}
\includegraphics[width=\columnwidth]{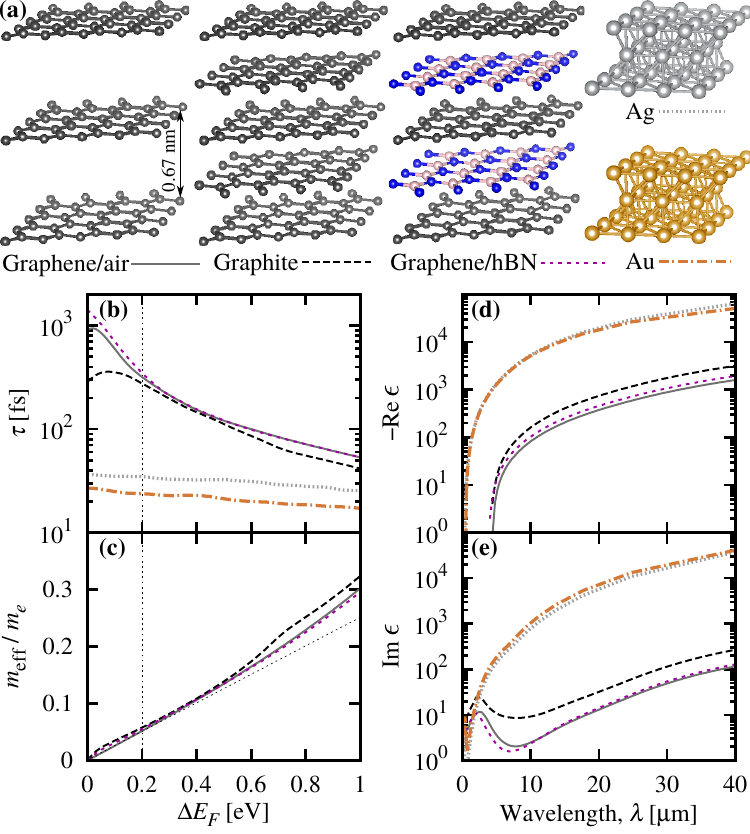}
\caption[Schematic, scattering time and dielectric functions]{(a) Schematic structures of 2D materials, graphene-derived vdW heterostructures
and 3D noble metals in which we investigate relaxation dynamics, dielectric properties and reflectivity.
(b) Comparison of the average lifetime $\tau$
as a function of the Fermi level in 2D materials, heterostructures and the best-case 3D metals: Ag and Au.
(c) The effective mass parameter for the stack, as a function of the E$_F$, connects
our analytical understanding of atomic-metamaterials with \emph{ab initio} calculations.
In panels (d) and (e) we show the $\Re(\epsilon)$ and $\Im(\epsilon)$
of graphene-based heterostructures in comparison with the permittivity of gold and 
silver. For wavelengths above $10~\mu$m, doped graphene-based heterostructures exhibit higher reflectance compared to Ag and Au. Results for the heterostructures are shown at 0.2eV Fermi level, as this is the doping regime where they reflect maximally for wavelengths above $10~\mu$m.
\label{fig:SchematicTauEpsilon}}
\end{figure}

While the electronic and phononic properties of graphene-based heterostructures have been widely investigated,
the photonic research on these materials has been focused on graphene layers
separated by hundreds of nanometers-microns,\cite{C4NR03143A} which is the regime
of optical metamaterials.\cite{Wang:2013yg, DaiS.:2015sf, Caldwell:2016rm}
In contrast, we focus on atomistic photonics where graphene (and hBN)
monolayers are at their equilibrium separation ($3.3 - 3.4$~\AA),
as shown in Fig.~\ref{fig:SchematicTauEpsilon}(a); results for the
free-standing graphene case with double the layer spacing are shown
throughout for comparison to highlight the effect of inter-layer interactions. These inter-layer electronic interactions, which we account for using ab-initio electronic calculations\cite{GraphiteHotCarriers}, make this regime distinct from conventional photonic crystals and metamaterials.

We use our previously established \emph{ab initio} methodology\cite{PhononAssisted,TAparameters}
to evaluate the net dielectric tensor $\bar{\epsilon}(\omega)$ of the graphene and adjacent dielectric layers (if any).
For the layered 2D materials, the dominant in-plane dielectric response is more conveniently expressed
as a sheet conductance via $\epsilon_\parallel(\omega) = 1 + 4\pi i\sigma(\omega)/(\omega d)$,
where $d$ is the spacing between layers (length of unit cell normal to the plane of layers).
Using this relation, we obtain \emph{ab initio} predictions of the sheet conductance,
\begin{equation}
\sigma(\omega) = \frac{\sigma_0}{1-i\omega\tau} + \sigma\sub{direct}(\omega).
\label{eqn:sigma}
\end{equation}
The first term captures the intraband response of free carriers in the material,
where $\sigma_0$ is the zero-frequency sheet conductance and $\tau$ is the (Drude)
momentum-relaxation time of the electrons, both of which we extract from a
linearized Boltzmann equation using an \emph{ab initio} collision integral
for electron-phonon scattering processes.\cite{PhononAssisted}
The second term, which describes the effect of interband transitions, is
evaluated using Fermi's Golden rule for the imaginary part,
and then the Kramers-Kronig relation for the real part.

All these calculations use density-functional theory predictions for the
energies and matrix elements of both the electrons and the phonons,
which automatically accounts for detailed electronic structure effects such as
inter-layer interactions and response of electrons far from the Dirac point,
as well as scattering against both acoustic and optical phonons
including Umklapp and inter-valley processes.
See Ref.~\citenum{PhononAssisted} for details on the theoretical framework
and the Methods section for computational details.
For graphene and its heterostructures, changing the Fermi level $E_F$,
changes the equilibrium electron occupation factors
in the Boltzmann equation as well as Fermi Golden rule.
This affects all of the quantities in (\ref{eqn:sigma}),
and we account for this by explicitly evaluating these quantities for several
different values of $E_F$ ranging from the neutral (undoped) value to 1~eV above it.

The intraband response (first term of (\ref{eqn:sigma})) dominates at low frequencies.
Before delving into the full optical response, we first examine the properties of this term intuitively.
The strength of the optical response is controlled by $\sigma_0$,
which within the approximate Drude model is $ne^2\tau/m\sub{eff}$
(used here only for the discussion in this paragraph),
where $n$ is the carrier density, $\tau$ is the relaxation time
and $m\sub{eff}$ is the effective mass.
Figure~\ref{fig:SchematicTauEpsilon}(b) and (c) respectively compare
$\tau$ and $m\sub{eff}$ of the graphene heterostructures and the noble metals,
as a function of the change in Fermi level $\Delta E_F$ from the neutral value
(Dirac point for the graphene cases).
We highlight the extremely large $\tau\sim 1$~ps for undoped graphene in air,
which drops to $\sim 200$~fs in graphite due to interlayer interactions.
In contrast, encapsulating graphene with boron nitride layers increases
the undoped $\tau$ even further to $\sim 2$~ps, despite having the same
spacing between adjacent 2D layers as graphite.
With increasing $E_F$ and carrier density, $\tau$ drops rapidly because of
increasing phase-space for electron-phonon scattering, which in turn is
because of increasing electronic density of states near the Fermi level, $g(E_F)$.
For comparison, we also show the relaxation times for noble metals,
gold and silver which are much smaller ($\sim 30$ and 40~fs respectively)
and mostly insensitive to $E_F$ since $g(E_F)$ depends weakly on $E_F$.
(In any case, $E_F$ cannot be changed easily for these metals in experiment.)
For the linear dispersion relation of graphene, the appropriate effective mass
for the Drude model is $m\sub{eff} = \Delta E_F/v_F^2$ (rather than
from band curvature, which would imply $m\sub{eff}\to\infty$).
This $m\sub{eff}$ increases linearly in the ideal case of perfect
linear dispersion, with slight deviations due to band structure effects,
but is consistently smaller than $m\sub{eff} \approx m_e$  of metals.
Therefore, the graphene based heterostructures have much higher $\sigma_0$
per carrier (mobility), due to a greater $\tau$ and a lower $m\sub{eff}$.
These properties of greater $\tau$ and lower $m\sub{eff}$ in graphene-based heterostructures 
compared to noble metals motivate us to investigate their optical properties.

Although the maximum electronic scattering time and minimum effective mass are seen at the charge neutrality point ($\Delta E_F\approx 0$~eV), graphene at 0~eV does not contain enough carriers to produce a strong metallic response. 
We find that $\Delta E_F\approx 0.2$~eV is the optimum doping level which
provides the best tradeoff between increasing carrier density and
correspondingly decreasing scattering time; we use the predicted
dielectric functions at this doping level for all remaining calculations.
Figure~\ref{fig:SchematicTauEpsilon}(d,e) compares the real and imaginary parts of the
in-plane permittivity of the graphene heterostructures and the noble metals.
The permittivities of the heterostructures in the infrared is smaller than
that of the noble metals by typically two orders of magnitude,
but additionally, the ratio of imaginary to real parts is
smaller by an extra factor of 2 - 3.
This smaller imaginary-to-real ratio compensates for the lower plasma frequency
and makes the graphene-based materials better reflectors in the IR
by reducing losses, as we discuss below. This also suggests a drastic reduction in plasmonic losses in comparison to Ag and Au, which results in improved reflective properties in the IR regime. 
We point out here that results pertaining to Ag and Au are also derived \emph{ab initio}, 
assuming perfect crystalline metals.

\begin{figure}
\includegraphics[width=\columnwidth]{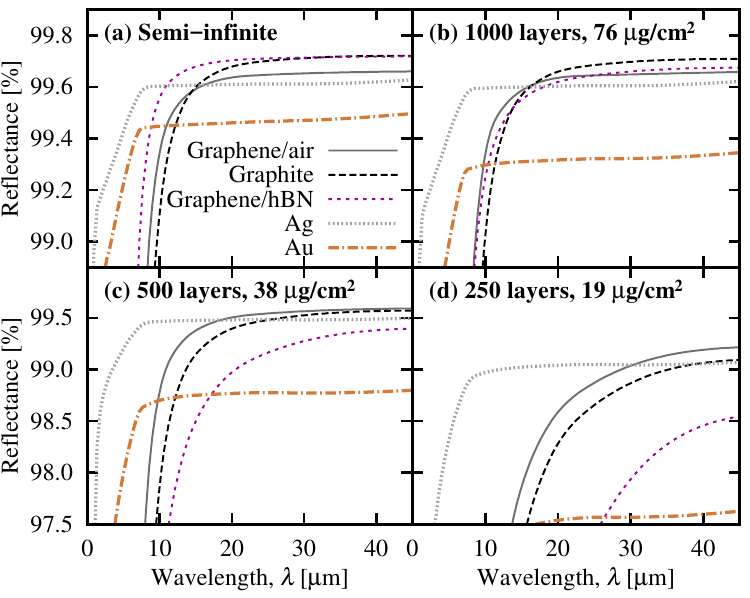}
\caption[Reflectance]{Normal incidence reflectance of graphene-air metamaterial, doped graphite, graphene-hBN heterostructure
(all doped at 0.2~eV) and comparison to Ag, and Au. Remarkably, the reflectance of graphene-based heterostructures surpasses
conventional noble metals in the mid-IR, above 10$\mu$m. Panels correspond to the same (per area) mass densities: (a): bulk, (b): 1000 layers,
corresponding to 72.4nm of Ag and 39.4nm of Au. (c): 500 layers, corresponding to 36.2nm of Ag and 19.7nm of Au.
(d): 250 layers, corresponding to 18.1nm of Ag and 9.85nm of Au. Note that the silver and gold values are based on \emph{ab initio} results for perfect conditions and therefore are the theoretical limit for these metals. Experimentally realized metallic thin films will compromise the magnitude of reflectance.
\label{fig:reflectance}}
\end{figure}

To realistically compare the utility of graphene vdW heterostructures as reflectors, we transition
from the infinite stacks of 2D layers considered above to stacks of finite thickness.
To do this, we perform electromagnetic transfer matrix calculations for layered media,\cite{YehPochi}
modified to account for the the non-vanishing surface current density in the 2D sheets\cite{PhysRevB.87.075416}.
Figure~\ref{fig:reflectance} compares the reflectance of these materials
for the bulk (semi-infinite) limit, as well as for finite stacks of
1000, 500 and 250 sheets with the Ag/Au slabs of the same (per area) mass density.
The reflectance of the vdW heterostructures surpasses that of the noble metals
above a critical wavelength, $\lambda \gtrsim 10\mu$m for the semi-infinite case,
$\gtrsim 15\mu$m for 1000 layers, $\gtrsim 20\mu$m for 500 layers and $\gtrsim 20\mu$m for 250 layers.
Note that silver and gold slabs of mass densities 19$\mu$g/cm$^2$ (that of 250 sheets)
correspond to thicknesses of 18.1nm and 9.85nm respectively; thin-film deposition of such ultra-thin
silver and gold films is known to lead to island formations and grain-boundary inhomogeneities.
In contrast, the van der Waals nature of the graphene heterostructures provides a pathway
to uniform, wafer-scale components with tremendous potential for ultra-light, long-IR mirrors.

\begin{figure}
\includegraphics[width=\columnwidth]{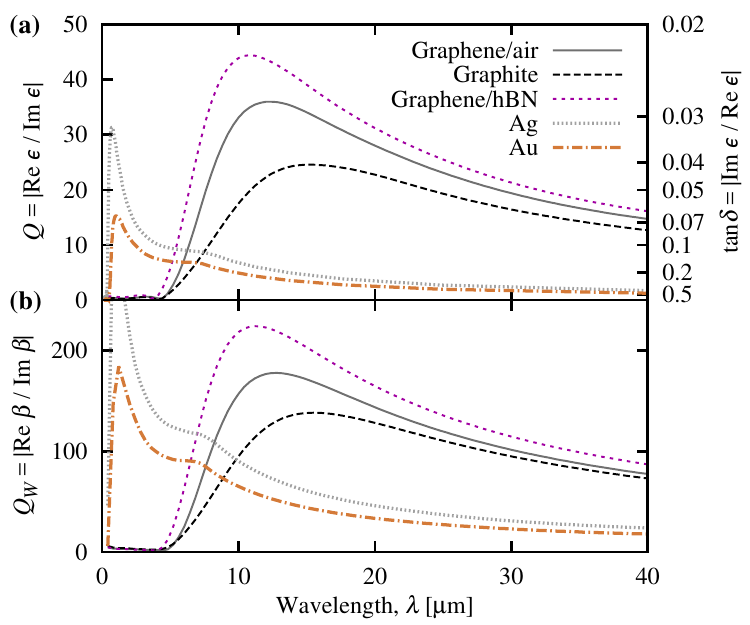}
\caption[Quality factors]{
(a) Material quality factor $Q=\Re\epsilon$/$\Im\epsilon$ and loss tangent($\tan\delta=Q^{-1}$)
and (b) the plasmonic modal quality factor $Q_W = \Re\beta$/$\Im\beta$ for the symmetric mode
in a metal-dielectric-metal waveguide, where the `metal' is either a conventional noble metal
or a vdW heterostructure and the dielectric is 0.1~$\mu$m of air.
\label{fig:Qfactors}}
\end{figure}

In discussing the permittivities shown in Fig.~\ref{fig:SchematicTauEpsilon}(d,e) above,
we noted that the ratio of imaginary to real permittivity (the loss tangent $\tan\delta$)
was smaller in the vdW heterostructures than in the noble metals, which resulted in lower losses.
A metric that directly illustrates the superior performance of graphene-heterostructures compared to noble metals is the material quality factor $Q \equiv 1/\tan\delta
\equiv \Re\epsilon$/$\Im\epsilon$ and its inverse, the material loss tangent $\tan\delta$, shown in Fig.~\ref{fig:Qfactors}(a). 
For the chosen doping level of 0.2~eV, and for wavelengths $\lambda > 15~\mu$m (mid-IR),
the vdW heterostructures are expected to show an order of magnitude higher performance
(higher $Q$-factor / lower loss tangent) as electromagnetic materials than noble metals.
The material Q-factor peaks below $\lambda \sim 1~\mu$m for noble metals and at $\lambda \sim 10~\mu$m for the vdW heterostructures corresponding to their respective plasma frequency regimes, thereby making these materials
particularly suited for mid-long wave IR applications.

\begin{figure}
\includegraphics[width=\columnwidth]{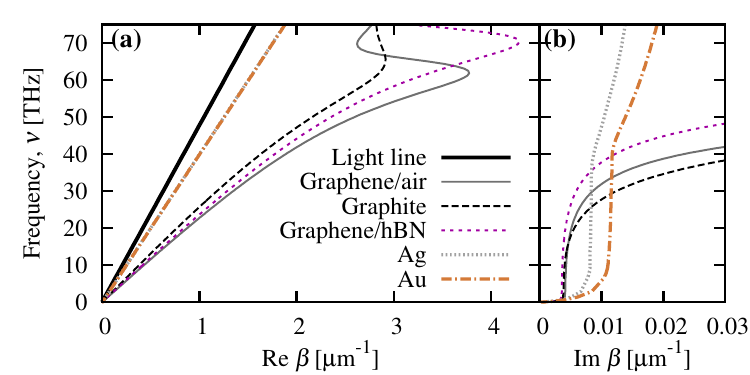}
\caption[Waveguide dispersion relations]{
Comparison of dispersion relations for the symmetric mode in the
metal/dielectric/metal waveguides considered in Fig.~\ref{fig:Qfactors}(b).
The larger $\Re(\beta)$ with vdW heterostructures corresponds
to smaller effective wavelength and improved mode confinement.
The smaller $\Im(\beta)$ for $\nu < 25$~THz corresponds to 
a two-fold increase in the decay length of the mode.
\label{fig:Dispersion}}
\end{figure}

Finally we investigate the performance of these materials
in a realistic geometry: a metal-dielectric-metal waveguide.
Such structures have been widely investigated in the visible part
of the spectrum using plasmonic metals such as Ag and Au as
the metal in the structure.\cite{Lezec430, PhysRevB.73.035407}
Based on our results, we envision replacing the metal with
a vdW heterostructure to improve performance for mid-IR applications.
For simplicity, we take the dielectric to be $0.1~\mu$m of air.
We use our predicted dielectric functions in the formalism from
Alu et. al \cite{Alu:06} to evaluate the plasmonic in-plane wavenumber ($\beta$).
Fig.~\ref{fig:Qfactors}(b) compares the plasmonic quality factor $Q_W$ for the
symmetric mode between vdW heterostructures and metals while in Figure~\ref{fig:Dispersion} we explicitly 
compare the corresponding dispersion relations.
The plasmonic quality factor $Q_W$ indicates the propagation length in number of mode-wavelengths which shows a three-five fold improvement for the vdW heterostructures compared to noble metals for $\lambda > 15~\mu$m.
In absolute terms, compared to propagation distances of 50-60 mode-wavelengths with use of Ag or Au in the mid-IR,
a graphene/hBN-based waveguide supports propagation distances that may exceed 200 mode-wavelengths.

Figure~\ref{fig:Dispersion} shows the dispersion relation of the symmetric waveguide mode, exhibiting the typical plasmonic asymptotic approach to the surface plasma frequency $\omega\sub{ps}$ \cite{Lezec430, PhysRevB.73.035407}.
The modal wavenumber ($\Re\beta$) reaches a maximum wavenumber at $\omega\sub{ps}$, and then returns towards the light line as losses increase and the mode becomes leaky.
This feature is observable in the mid-IR frequency range for all the vdW-heterostructures,
with slightly different resonance frequencies ranging from 60-70~THz, while corresponding
features for the noble metal waveguides will appear above 500~THz in the visible-ultraviolet regime.
The larger in-plane wavenumbers ($\Re\beta$) of the modes in the vdW-heterostructure-based waveguide
indicate shorter in-plane wavelength and correspondingly higher mode-confinement in perpendicular directions.
Most importantly, the smaller imaginary parts ($\Im\beta$) illustrate larger propagation distances, as shown in Fig.~\ref{fig:Qfactors}(b).

We have shown using a combination of \emph{ab initio} methods and optical transfer matrix calculations
that long electron relaxation times in graphene-based heterostructures leads to improved
optical properties compared to noble metals especially in the mid-infrared regime.
In particular, we predict an order of magnitude improvement in plasmonic material performance, and reflectivity exceeding that of Ag and Au for materials with substantially reduced mass density. 
This suggests the possibility of replacing current noble-metal components in optoelectronic devices
with 2D van der Waals heterostructures, which can also be tuned in real time, for improved performance in the mid-IR frequency range
in active waveguiding systems, Salisbury screens for perfect absorption,
for engineering Purcell enhancements, and especially in aerospace applications,
where mass-density becomes an important figure of merit.
The increased carrier density in graphene-based vdW heterostructures required for unlocking their low-loss plasmonic response in the mid-IR may be achieved by gating alternating graphene layers separated by hBN. The doping of graphite, for example with Li intercalation is also an option for experimental realization of the concept we are proposing here.

\section*{Methods: Computational details}
\label{sec:CompDetails}

We use \emph{ab initio} density-functional theory (DFT) calculations
of the electronic band structure, phonon dispersion relations and
electron-phonon matrix elements, interpolated using maximally-localized
Wannier functions\cite{MLWFmetal} for efficient and accurate Brillouin-zone
integration in the Boltzmann equation and Fermi Golden rule calculations.
For the DFT calculations, we use the `PBE' exchange-correlation functional,\cite{PBE}
a plane-wave basis at a kinetic energy cutoff of 30~Hartrees with the truncated
Coulomb potential approach for non-periodic geometries,\cite{TruncatedEXX}
and norm-conserving pseudopotentials,\cite{SG15}
all using the open-source JDFTx software.\cite{JDFTx}
See Ref.~\citenum{PhononAssisted} for theoretical and implementation details,
and Refs.~\citenum{GraphiteHotCarriers} and \citenum{PhononAssisted} respectively
for the computational details for the graphene-based materials and noble metals.

We utilize the \emph{ab initio}-derived 2D conductivities for the graphene/graphite/graphene-hBN heterostructure unit cells to 
perform transfer matrix-based electromagnetic calculations of reflectance/transmittance based on a transfer matrix formalism. 
We utilize those scattering parameters in a rigorous S-matrix based parameter retrieval for layered media as shown in Ref~\citenum{PhysRevB.91.155406}
to retrieve the effective permittivity of heterostructures composed of 1000, 500 and 250 sheets. 
With increasing thickness, the effective dielectric permittivity of finite stacks approaches
the bulk dielectric response shown in Fig.~\ref{fig:SchematicTauEpsilon}(d, e), as expected.

\section*{Acknowledgements}
GTP acknowledges fruitful discussions with Prof. P. Yeh. 
We also thank Prof. Harry A. Atwater, Prof. John D. Joannopoulous, Yi Yang and Tena Dubcek for 
helpful discussions. We acknowledge financial support from NG Next at the Northrop Grumman Corporation.
GTP acknowledges financial support from the American Association of University Women (AAUW).
PN acknowledges support from the Harvard University Center for the Environment (HUCE).
RS acknowledges startup funding from Rensselaer Polytechnic Institute (RPI). NE acknowledges partial support from the U.S. Air Force Office of Scientific Research Multidisciplinary University Research Initiative grant number FA9550-17-1-0002.
This research was supported (in part) by the U.S. Army Research Office under contract W911NF-13-D-0001.
Calculations in this work were performed on the BlueGene/Q supercomputer in the
Center for Computational Innovations (CCI) at RPI,
as well as in National Energy Research Scientific Computing Center,
a DOE Office of Science User Facility supported by the Office of Science
of the U.S. Department of Energy under Contract No. DE-AC02-05CH11231.
M.S. (reading and analysis of the manuscript) was supported by S3TEC, an Energy Frontier Research Center funded by the U.S. Department of Energy under grant no. DE-SC0001299.

\makeatletter \renewcommand{\@biblabel}[1]{#1.} \makeatother
\renewcommand{\bibpreamble}{\textbf{\large\uppercase{References}}}
\bibliographystyle{achemso}
\makeatletter{}\providecommand{\latin}[1]{#1}
\providecommand*\mcitethebibliography{\thebibliography}
\csname @ifundefined\endcsname{endmcitethebibliography}
  {\let\endmcitethebibliography\endthebibliography}{}

\end{document}